\documentclass[twocolumn,aps,psfig,epsfig,bm,showpacs,superscriptaddress]{revtex4} 
\usepackage{ulem}
\usepackage{epsfig}

\begin{document} 

\newcommand{\vk}{{\vec k}} 
\newcommand{\vK}{{\vec K}}  
\newcommand{\vb}{{\vec b}}  
\newcommand{\vp}{{\vec p}}  
\newcommand{\vq}{{\vec q}}  
\newcommand{\vQ}{{\vec Q}} 
\newcommand{\vx}{{\vec x}} 
\newcommand{\vh}{{\hat{v}}} 
\newcommand{\tr}{{{\rm Tr}}}  
\newcommand{\be}{\begin{equation}} 
\newcommand{\ee}{\end{equation}}  
\newcommand{\half}{{\textstyle\frac{1}{2}}}  
\newcommand{\gton}{\stackrel{>}{\sim}} 
\newcommand{\lton}{\mathrel{\lower.9ex \hbox{$\stackrel{\displaystyle 
<}{\sim}$}}}  
\newcommand{\ben}{\begin{enumerate}}  
\newcommand{\een}{\end{enumerate}} 
\newcommand{\bit}{\begin{itemize}}  
\newcommand{\eit}{\end{itemize}} 
\newcommand{\bc}{\begin{center}}  
\newcommand{\ec}{\end{center}} 
\newcommand{\bea}{\begin{eqnarray}}  
\newcommand{\eea}{\end{eqnarray}}

\newcommand{\comment}[1]{}
\newcommand{\ttbs}{\char'134}
\newcommand{\srt}{\mbox{$\sqrt{s}$}}
\def\GeV{{\rm GeV}}
\def\AGeV{{\rm A GeV}}
\def\GeVc{{\rm GeV/c}}
\def\AGeVc{{\rm A GeV/c}}
\def\MeV{{\rm MeV}}
\def\fm{{\rm fm}}
\newcommand{\bold}[1]{\mbox{\boldmath $#1$}}    
\newcommand{\pp}{\bold p}
\newcommand{\Bbar}{\mbox{$\bar{\rm B}$}}
\newcommand{\etal}{{\it et al.}}
\newcommand{\gsim}{\mbox{\raisebox{-0.6ex}{$\stackrel{>}{\sim}$}}\:}
\newcommand{\lsim}{\mbox{\raisebox{-0.6ex}{$\stackrel{<}{\sim}$}}\:}

\title{Influence of Bottom Quark Jet Quenching 
on  Single Electron  Tomography of Au+Au
}
 
\date{\today}
 
\author{Magdalena Djordjevic}
\affiliation{Department of Physics, Columbia University, 
             538 West 120-th Street, New York, NY 10027}
\author{Miklos Gyulassy}
\affiliation{Department of Physics, Columbia University, 
             538 West 120-th Street, New York, NY 10027}

\author{Ramona Vogt}
\affiliation{Nuclear Science Division, LBNL, Berkeley, CA 94720 and 
Physics Department, University of California, Davis, California 95616}

\author{Simon Wicks}
\affiliation{Department of Physics, Columbia University, 
             538 West 120-th Street, New York, NY 10027}

\begin{abstract} 
High transverse momentum single (non-photonic)
electrons are shown to be sensitive
to the 
stopping power  of both bottom, $b$, and charm, $c$, 
quarks in $AA$
collisions. 
We apply 
the DGLV theory of radiative energy loss to predict 
$c$ and $b$ quark jet quenching and compare the FONLL and PYTHIA
heavy flavor fragmentation and decay schemes.
We show that single electrons in the $p_T=5-10$ GeV
range 
are dominated by the decay of $b$ quarks rather than 
the more strongly quenched 
$c$ quarks in Au+Au
collisions at $\sqrt{s}=200$ $A$GeV.
The {smaller} $b$ quark energy loss, 
even for extreme opacities with gluon rapidity densities
up to 3500,
is predicted
to limit the nuclear modification factor, $R_{AA}$, of single electrons
to the range $R_{AA} \sim 0.5-0.6$, in contrast to
previous predictions of $R_{AA}\lton 0.2-0.3$ based on 
taking only $c$ quark jet fragmentation into account.
\end{abstract}


\maketitle 

{\em Introduction.} 
 
Recent data~\cite{WhitePapers} from the Relativistic Heavy Ion Collider (RHIC)
on ``perfect fluidity'' ~\cite{BNL}-\cite{Hirano:2005wx} and 
light quark and gluon jet quenching~\cite{Gyulassy:2003mc}-\cite{
Vitev:2002pf} 
provide direct evidence 
that a  novel form of strongly interacting  Quark Gluon Plasma (sQGP)
is created in central 
Au+Au collisions
at $\sqrt{s} = 200$ $A$GeV \cite{Gyulassy:2004zy}. 

In the near future, measurements of 
heavy quark jet quenching will provide further important
tests of
the transport properties of this new form of matter.
In particular, rare heavy quark jets are valuable independent probes of
the intensity of color field fluctuations in the sQGP 
because their high mass 
{($m_c \approx 1.2$ GeV, $m_b \approx 4.75$ GeV)
changes the sensitivity of both 
elastic and inelastic energy loss mechanisms
in a well defined way \cite{Dead-cone}-\cite{Armesto:2005iq}
relative to those of light quark and gluon 
jets  \cite{Gyulassy:2003mc}-\cite{Vitev:2002pf}.
Open heavy quark meson {($D$, $B$)} tomography
also has the unique advantage that - unlike light hadron ($\pi$, $K$) 
tomography that is sensitive to the large difference between 
quark and gluon energy loss -  
gluon jet {fragmentation into $D$ and $B$
mesons} can be safely neglected.

The ``fragility'' of light hadron tomography pointed out in 
{Ref.~\cite{Eskola:2004cr} is primarily due to the significant reduction
in} sensitivity of the attenuation pattern
to the sQGP density when the gluon jets originating from the interior
are too strongly
quenched. In that case, the attenuation of light hadrons
becomes sensitive to geometric
fluctuations of the jet production points near the surface ``corona''.

Heavy quarks, especially $b$ quarks,
are predicted to be significantly less fragile in the 
DGLV \cite{Djordjevic:2003zk}-\cite{Djordjevic:2004nq}
theory {of} radiative energy loss because {their energy loss is
expected to be considerably
smaller.
If radiative energy loss is the dominant jet 
quenching mechanism in the $p_T\sim 10 $ GeV
region, then heavy meson tomography 
could be a more sensitive tomographic probe of
the absolute scale of 
density evolution and the opacity of the produced sQGP.

However, one disadvantage
of heavy meson tomography
is that direct measurements
of identified high $p_T$
$D$ and $B$ mesons are very difficult with current
detectors and RHIC {luminosities} \cite{Harris:2005gn}.
Therefore, the first experimental studies of heavy quark attenuation
at RHIC have focused on
the attenuation of their single (non-photonic) 
electron decay products \cite{Adcox:2002cg}-\cite{Adler:2005ab}.

Some preliminary data \cite{Laue:2004tf}-\cite{Averbeck:2005mj}
{surprisingly suggest} 
that single electrons with $p_T\sim 5$ GeV
may {experience elliptic flow and
suppression patterns similar to light partons.} 
We emphasize in this letter that either 
result would have even greater implications than previously
thought about  the nature of the produced sQGP. 
If confirmed 
in the final analysis, the sQGP would have to be completely opaque
to even {$b$ quark} jets of $p_T
\sim 10 $ GeV, in
contradiction to
all radiative energy loss estimates so far. 

A significant complication of the heavy quark decay
lepton measurements is that 
estimates in Refs.~\cite{Lin:2004dk,Dong:2005} indicated
that bottom decay leptons may in fact 
dominate electrons from charm for $p_T > 3$ GeV
in $pp$ collisions.  
In this letter, we show that jet quenching further amplifies
the $b$ contribution to the lepton spectrum and 
strongly limits the nuclear modification factor of electrons
in $AA$ collisions.

The preliminary electron
data \cite{Laue:2004tf}-\cite{Averbeck:2005mj}
are so surprising that novel jet energy loss mechanisms 
may have to be 
postulated \cite{Molnar:2004ph}-\cite{Moore:2004tg}.
The elliptic flow of high $p_T$ heavy quarks
can be accounted for, e.g., if the {\it elastic}
cross sections of all partons, including bottom, are assumed to be 
anomalously
enhanced to $>20$ mb, far in excess of perturbative QCD 
predictions, up to at least $p_T \sim 10$ GeV.
While {these enhanced cross sections} could lead to 
{heavy flavor elliptic flow
at the pion level} even at high $p_T$, 
{they may greatly overestimate the 
attenuation of light 
and heavy flavored hadrons} \cite{Moore:2004tg}-\cite{HorowitzGM}.

{Given the critical role that single electron tomography of the sQGP
may play in the near future, it is especially 
important to scrutinize the
theoretical uncertainties and robustness of current predictions.
This is the aim of this letter.}

{\em Theoretical framework.}\\
The calculation of the lepton spectrum includes initial heavy quark 
distributions from perturbative QCD,
heavy flavor energy loss, heavy quark fragmentation into heavy hadrons, 
$H_Q$, and $H_Q$ decays
to leptons. The cross section is schematically written as: 
\begin{eqnarray}
\frac{E d^3\sigma(e)}{dp^3} &=& \frac{E_i d^3\sigma(Q)}{dp^3_i}
 \otimes
{P(E_i \rightarrow E_f )}\nonumber \\
&\otimes& D(Q \to H_Q) \otimes f(H_Q \to e) \nonumber \; ,
\end{eqnarray}
where $\otimes$ is a generic convolution. The electron decay
spectrum, $f(H_Q \to e)$, includes the branching ratio to electrons.
The change in the initial heavy flavor spectra due to energy loss is denoted
$P(E_i \rightarrow E_f)$.}

{The initial heavy quark $p_T$ distributions are computed at 
next-to-leading order with the code used in 
{Ref.~\cite{Cacciari:2005rk,MNR}}.
We assume the same mass and factorization scales as in Ref.~\cite{Vogt},
employing the CTEQ6M parton densities {\cite{cteq6m}}  with
no intrinsic $k_T$.}

As in {Ref.~}\cite{Djordjevic:2004nq},
we compute {heavy flavor} suppression with
the DGLV generalization~\cite{Djordjevic:2003zk} of the GLV opacity 
expansion~\cite{Gyulassy:2000er} to heavy quarks. We take into account 
multi-gluon fluctuations as in Ref.~\cite{GLV_suppress}.

The fragmentation functions $D(c\to D)$ and $D(b\to B)$, where $D$ and $B$
indicate a generic admixture of charm and bottom hadrons, are consistently
extracted from $e^+e^-$ data
\cite{Cacciari:2002pa,Cacciari:2003zu,Cacciari:2003uh}. The
charm fragmentation function~\cite{Cacciari:2003zu} depends on the parameter
$r$~\cite{Braaten:1994bz}.  We take $r = 0.04$ for $m_c = 1.2$ GeV. 
Bottom fragmentation instead depends on the  parameter $\alpha$
\cite{Kartvelishvili:1977pi} with
$\alpha = 29.1$ for $m_b = 4.75$~GeV.
The fragmentation is done by
rescaling the quark three-momentum at a constant angle in the laboratory frame.

The leptonic decays of $D$ and $B$ mesons are 
controlled by measured decay spectra and branching ratios.
The spectrum for primary $B\to e$ decays has been measured recently 
\cite{Aubert:2004td,Mahmood:2004kq}. The fit to this data 
{\cite{Cacciari:2005rk}} is assumed to
be valid for all  bottom hadrons.  Preliminary CLEO data on the inclusive 
semi-leptonic electron 
spectrum from $D$
decays \cite{yelton} have also been fitted {\cite{Cacciari:2005rk}} 
and assumed
to be identical for all charm hadrons. The contribution of leptons from
secondary $B$ decays $B\to D\to e$ is obtained
as a  convolution of the $D\to e$ spectrum
with a parton-model prediction for $b\to c$ decay 
{\cite{Cacciari:2005rk}}.
The resulting electron spectrum is very soft, making it a negligible 
contribution to the total, particularly at $p_T > 2$ GeV.
The appropriate effective branching ratios are~\cite{Eidelman:2004wy}:
$B(B\to e) = 10.86 \pm 0.35$\%, $B(D\to e) = 10.3 \pm 1.2$\%,
and $B(B\to D\to e) = 9.6 \pm 0.6$\%.

{
The uncertainty in our results due to the choice of fragmentation and decay 
schemes is studied using the corresponding PYTHIA {\cite{Sjostrand:2000wi}}
routines, assuming
Peterson fragmentation {\cite{Peterson}} with a range of parameters.
}

\begin{figure}[ptDist] 
\epsfig{file=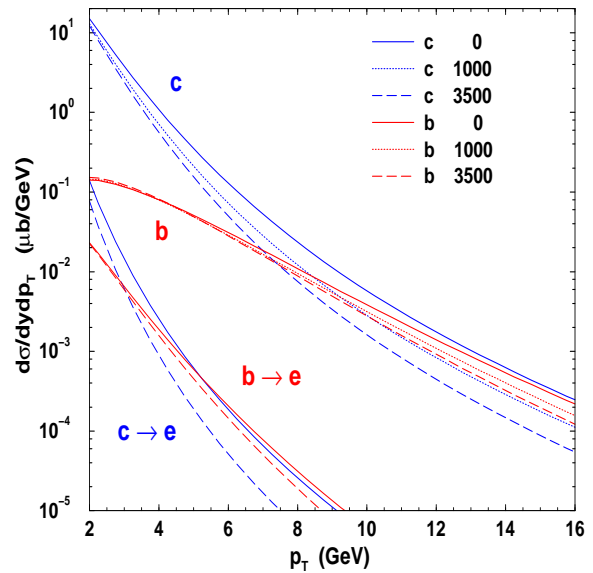,height=3.0in,width=3.0in,clip=5,angle=-90} 
\begin{minipage}[t]{8.6cm}  
\caption{\label{fig:ptDist} The differential cross section (per nucleon pair)
of charm (upper blue) and bottom (upper red) quarks 
calculated to NLO in QCD
{\protect~\cite{Cacciari:2005rk}}
compared to single electron distributions 
{calculated with the fragmentation and decay scheme of
Ref.~{\protect \cite{Cacciari:2005rk}}.}
The solid, dotted and long dashed curves show the
effect of DGLV heavy quark
quenching with initial rapidity densities of $dN_g/dy=0,1000,\; 
{\rm{and}}\; 3500$, respectively.}
\end{minipage} 
\end{figure} 


\begin{figure*}[bht] 
\epsfig{file=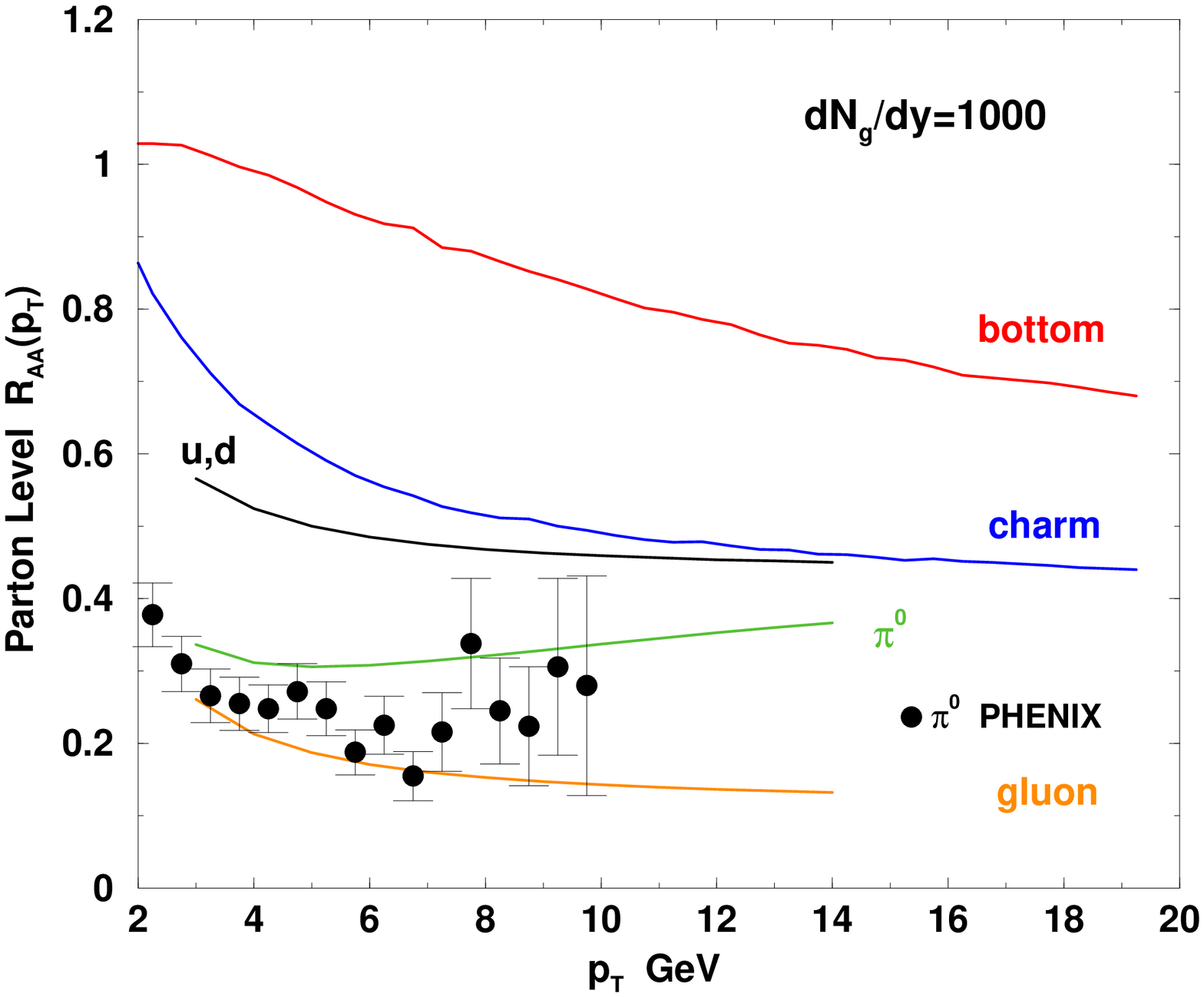,height=2.8in,width=2.8in,clip=5,angle=-90} 
\epsfig{file=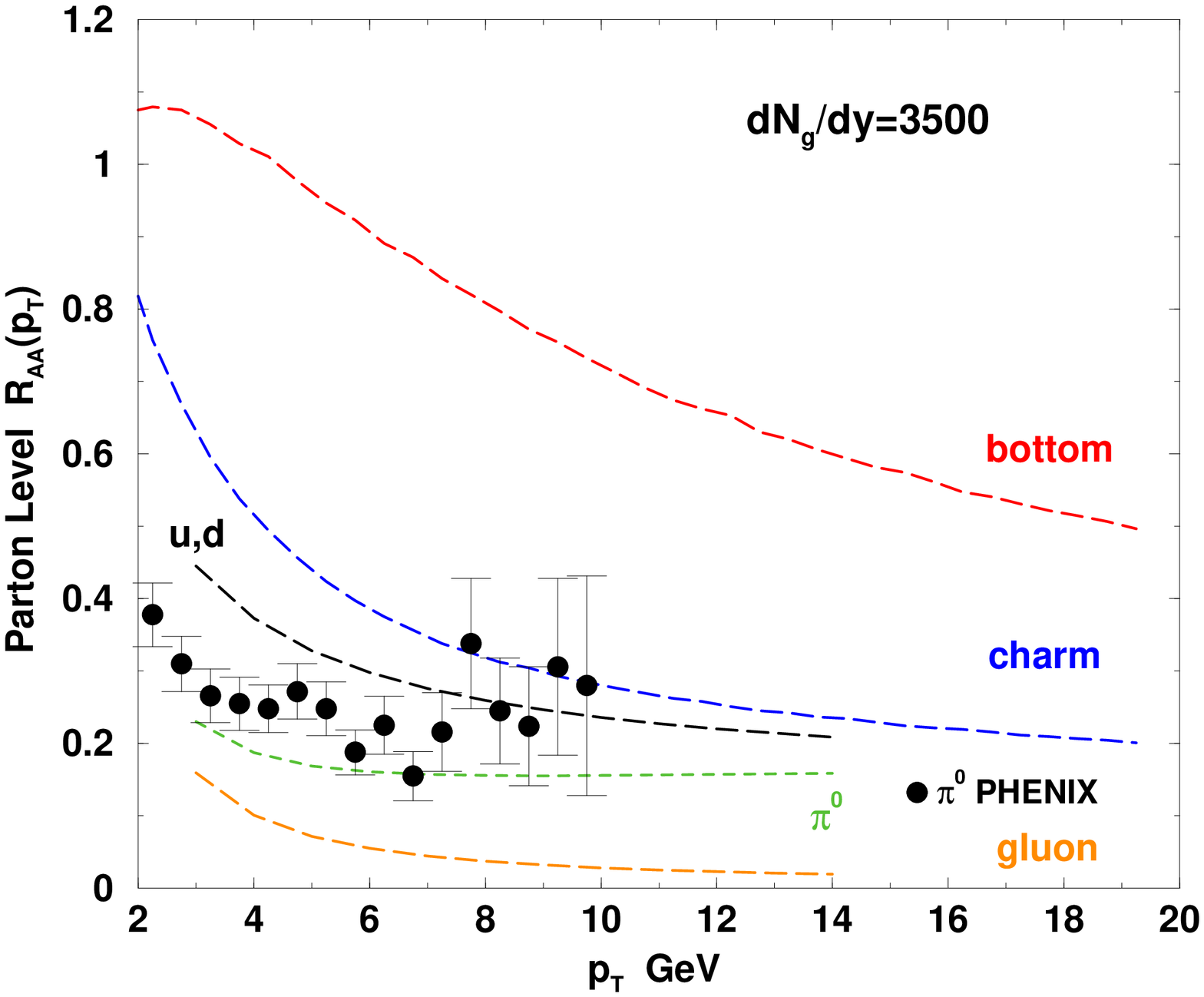,height=2.8in,width=2.8in,clip=5,angle=-90} 
\begin{minipage}[t]{15.cm}  
\caption{\label{fig:qRAA} Heavy quark jet quenching before
fragmentation into mesons for $dN_g/dy=1000$ (left) and 3500
(right) are compared to light ($u$, $d$) quark and gluon 
quenching. The resulting $\pi^0$ $R_{AA}$ is compared to the central 0-10\%
PHENIX data  {\protect\cite{Adler:2003qi}}. }
\end{minipage}
\end{figure*}

\begin{figure*}[bht] 
\epsfig{file=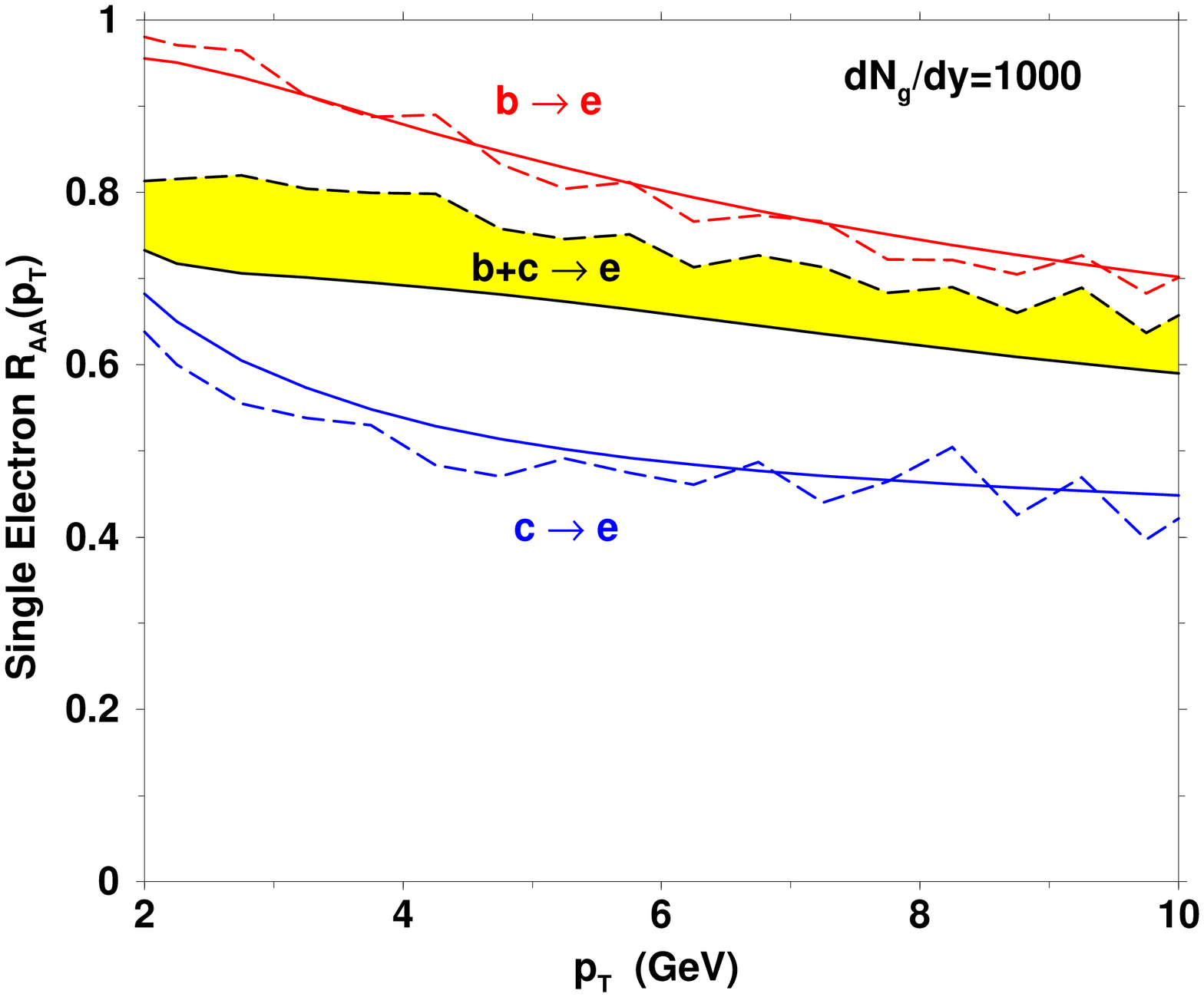,height=2.8in,width=2.8in,clip=5,angle=-90} 
\epsfig{file=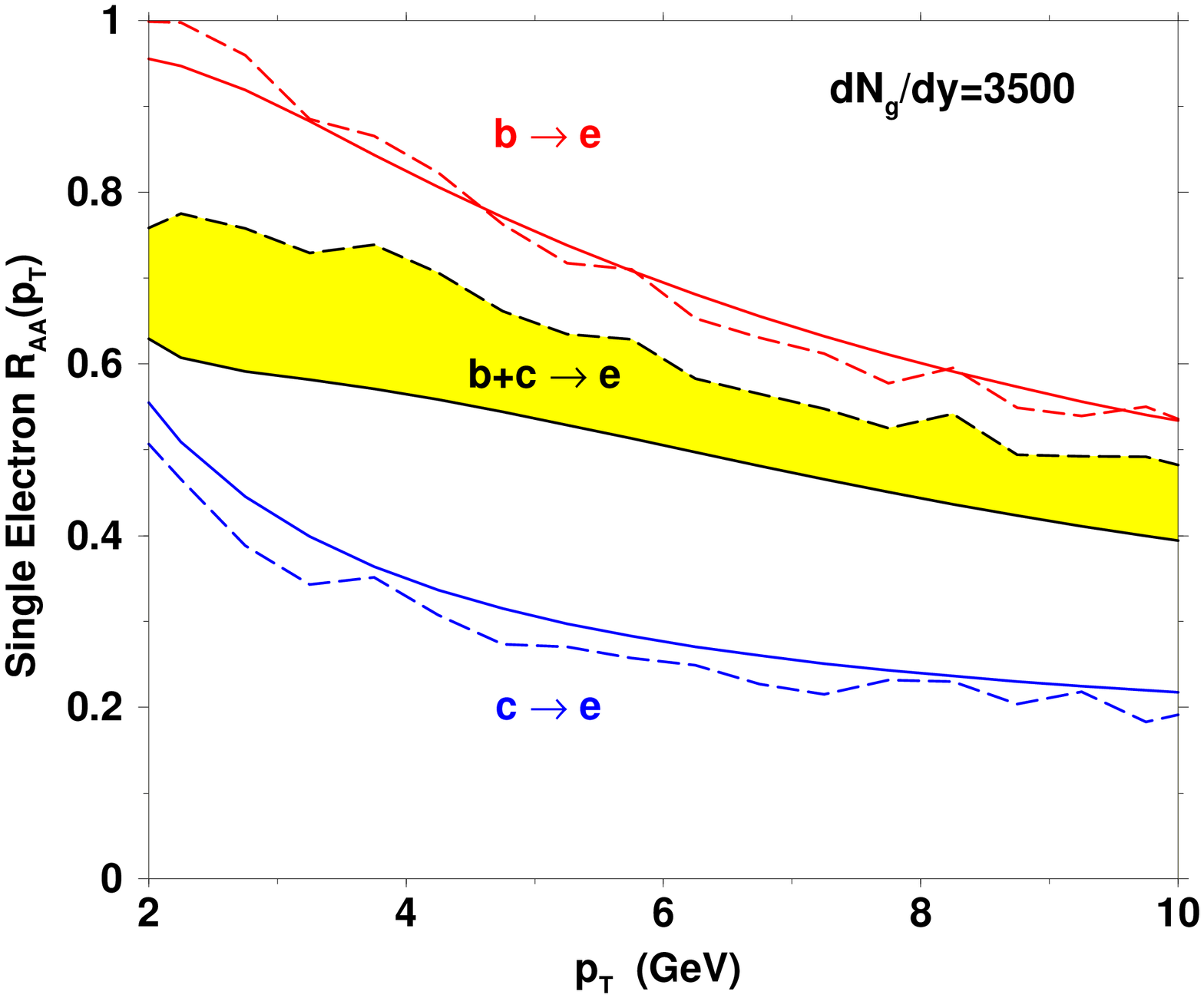,height=2.8in,width=2.8in,clip=5,angle=-90} 
\begin{minipage}[t]{15.cm}  
\caption{\label{fig:eRAA} Single electron attenuation pattern for 
initial $dN_g/dy=1000$, left,  and $dN_g/dy=3500$, right.
The solid curves employ 
the fragmentation scheme and lepton decay parameterizations of 
Ref.~{\protect \cite{Cacciari:2005rk}}
while the dashed curves use the 
{Peterson function with $\epsilon_c = 0.06$ and $\epsilon_b = 0.006$} and the
decay to leptons employed by the PYTHIA Monte Carlo.
Note that even for the extreme opacity
case on  the right the less quenched 
$b$ quark jets dilute $R_{AA}$ so much that
the modification of the
combined electron yield from both $c$ and $b$ jets does not 
fall below
$\sim 0.5-0.6$ near $p_T\sim 5$ GeV. }
\end{minipage}
\end{figure*}

To compute the medium induced 
gluon radiation spectrum, we need to include in general three 
 effects: 1)~the 
Ter-Mikayelian or massive gluon effect~\cite{DG_TM, DG_PLB}, 
2)~transition 
radiation~\cite{Zakharov} 
and 3)~medium-induced energy loss~\cite{DG_PLB, Djordjevic:2003zk}.
In Ref.~\cite{DG_Trans_Rad}, it was shown
that first two effects nearly cancel and can thus be neglected for
heavy quark suppression at zeroth order in opacity. 
We therefore only compute the medium-induced
gluon radiation spectrum \cite{Djordjevic:2003zk}.
We employ the effective static medium approximation formula
\bea 
\frac{ d N_{\rm ind}^{(1)}}{d x} &=&  \frac{C_{F}\alpha_s}{\pi} 
\frac{L}{\lambda_g} \int_0^\infty 
\frac{ 2 \mathbf{q}^2 \mu^2 d\mathbf{q}^2}{( \frac{4 E x}{L} )^{2} 
+ (\mathbf{q}^{2} + m^{2}x^{2} + m_{g}^{2})^{2}} 
\nonumber \\
&\times& \int \frac{ d\mathbf{k}^2 \; \theta (2 x (1-x) p_T-
|\mathbf{k}|)} {(( |\mathbf{k}|-|\mathbf{q}|)^{2} + \mu^{2})^{3/2} 
(( |\mathbf{k}|+|\mathbf{q}|)^{2} + \mu^{2})^{3/2}} 
\nonumber \\
&\times&\left\{ \mu^2+ (\mathbf{k}^{2}-\mathbf{q}^{2}) 
\frac{\mathbf{k}^{2} - m^{2}x^{2} - m_{g}^{2}}
{\mathbf{k}^{2} + m^{2}x^{2} + m_{g}^{2}} \right\} \, \, . 
\label{gluon_rad}
\eea

{Here $E = \sqrt{p_T^2 + m^2}$ is
the initial energy of a heavy quark of mass $m$,}
$\mathbf{k}$ is the transverse momentum of the radiated gluon and 
$\mathbf{q}$ is the momentum transfer to the jet. The opacity
of the medium to {\it radiated gluons} is
$L/\lambda_g={9\pi\alpha_s^2  }/{2}\int d\tau {\rho(\tau)}/{\mu^2(\tau)}$ 
where $\mu \approx g (\rho/2)^{1/3}$ is local Debye mass in a perturbative 
QGP. The gluon density at proper time $\tau$ 
is related to the initial
rapidity density of the produced gluons by
$\rho(\tau)\approx {dN_{g}}/{dy \tau \pi R^2} $ with 
$R=6$~fm 
in central collisions assuming a uniform
cylinder undergoing a Bjorken 1+1D expansion. 
Transverse expansion does not significantly affect
the integrated energy loss \cite{Gyulassy:2001kr}.

The Ter-Mikayelian effect at first order
in opacity is due to a
asymptotic transverse gluon mass in the medium,
$m_{g}\approx \mu / \sqrt{2}$. We assume 
$\alpha_s=0.3$. 
The induced radiative energy loss fluctuation
spectrum, $P(E_i\rightarrow E_f)$, was computed as in
{Ref.}~\cite{GLV_suppress}{,} starting from
the average induced gluon spectrum
in the effective {\it static} medium approximation
given by Eq~(\ref{gluon_rad}).
In this approximation {the} effective static $\rho$ 
is approximated by $\rho(\langle \tau \rangle)$ 
with {$\langle \tau \rangle = R/2=3$ fm} and $L=R$. 
We have checked
that {the more 
numerically intensive Bjorken expansion gives very similar results.}

{Note that $k_{\rm max} = 2x (1-x)p_T$
in Eq.~(1) instead of $k_{\rm max} = xE$, as in Ref.~\cite{Djordjevic:2003zk}.
There is a 20\% theoretical uncertainty in $R_{AA}$ due to the range of
reasonable kinematic bounds.}

{\em Bottom versus Charm  quark suppression} 

Figure~\ref{fig:ptDist} shows the $c$ and $b$ quark distributions
at midrapidity before fragmentation. The solid curves indicate that,
at NLO, $b$ production becomes comparable to $c$ 
production in the vacuum 
only for $p_T\gton 15$ GeV. However, jet quenching is greater
for the lighter $c$ quark, and for the default gluon density, $dN_g/dy=1000$
\cite{Vitev:2002pf},
the more weakly quenched $b$'s dominate over the more strongly quenched
$c$'s for $p_T\gton 9$ GeV. For more  extreme opacities,
characterized here by $dN_g/dy=3500$, the cross over shifts down to
$p_T\approx 7$ GeV. With the fragmentation and decay scheme of
Ref.~\cite{Cacciari:2005rk}, the electron decay distributions, 
$c\rightarrow e$ and $b\rightarrow e$, are seen
to cross each other at $p_T\sim 5.5$ GeV when the $c$ and $b$ quarks are
not quenched, reduced to $p_T\sim 3$ GeV for $dN_g/dy=3500$.
The electron results for $dN_g/dy=1000$, lying between the solid 
and long-dashed 
curves in Fig.~\ref{fig:ptDist}, are not shown for clarity.
Thus electrons in the $p_T\sim 5$ GeV region are 
sensitive to $b$ and $c$ quark quenching.

The parton level quenching is shown in detail in Fig.~\ref{fig:qRAA}
by the nuclear modification factor, $R_{AA}(Q)= dN_Q(p_T,dN_g/dy)/ 
dN_Q(p_T,0)$ with $Q=g,u,d,c$ and $b$. 
The left-hand side shows results for the
default case, $dN_g/dy=1000$ \cite{
Vitev:2002pf} , while the right-hand side shows the high opacity
case, $dN_g/dy=3500$. For comparison, we also show the 
PHENIX \cite{Adler:2003qi} data on the $\pi^0$ nuclear modification factor
measured in  the central 0-10\% of
Au+Au collisions at $\sqrt{s}=200$ $A$GeV. As expected, 
gluon quenching is largest due to its color Casimir factor and its small
in-medium mass.  The ``dead cone effect'' \cite{Dead-cone} is
seen by comparing $c$ quark
quenching to light $u,d$ quenching at $p_T<10$ GeV.
For $p_T>10$ GeV $\gg m_c$, the mass difference between the charm and light
quarks is almost negligible \cite{DG_Trans_Rad}.

However, in both cases, $b$ quark quenching
remains significantly smaller 
than that of the light and charm quarks for 
$p_T \lton 20$ GeV since $p_T/m_b$ is not large. The effect of the $b$ mass 
can therefore never be neglected
in the RHIC kinematic range.

{Figure~\ref{fig:qRAA} also shows an
estimate of $\pi^0$ quenching assuming}
\begin{equation}
R_{AA}(\pi^0)\approx  f_g\; R_{AA}(g) + (1-f_g)\; R_{AA}(u)
\;\; ,
\label{pi0}\end{equation}
where $f_g\approx \exp[-p_T/10.5\;{\rm GeV}]$ is the fraction
of pions with a given $p_T$ that arise from gluon jet fragmentation. 
The approximate form is a fit to a leading order QCD
calculation at $\sqrt{s}=200$ $A$GeV, discussed in 
Refs.~\cite{Vitev:2004gn,Adil:2004cn}.
The approximation in Eq.~(\ref{pi0}) is strictly valid only
for pure power law gluon and quark distributions with a
$p_T$-independent spectral index. However, it provides
a simple estimate that shows that $\pi^0$ quenching is primarily controlled
by  light quark quenching above 10 GeV. In addition,
Fig.~\ref{fig:qRAA} shows that current data would be incompatible
with radiative $g$, $u$ and $d$ quenching if the medium
had an opacity greater than that of the $dN_g/dy=3500$ case
considered on the right-hand side.

We note that the $c$ quark quenching predicted 
in Fig.~\ref{fig:qRAA} with $1000\leq dN_g/dy \leq 3500$ 
is similar to the quenching range
predicted in Fig.~2 of Ref.~\cite{Armesto:2005iq}
for the effective transport coefficient $\hat{q}=\mu^2/\lambda_g$
in the range $4 \leq \hat{q} \leq 14$ GeV$^2$/fm. 
For a $c$ quark with $p_T \sim 12$ GeV, for example,
we predict $R_{AA}(c)\approx 0.25-0.5$ in this range, as does
Ref.~\cite{Armesto:2005iq} for the same factor 
of 3.5 variation of the sQGP density.

Our primary new observation is that since $b$ quark quenching
is greatly reduced relative to $c$ quenching, if heavy quark tomography
is performed via single electron suppression patterns,
the lower $b$ quenching strongly limits the possible electron quenching,
as we show in Fig.~\ref{fig:eRAA}. For electrons arising from
$c$ fragmentation and decay, we again confirm the predictions of
Ref.~\cite{Armesto:2005iq}. However, for electrons
arising from $b$ decay, there is only a modest amount
of quenching. Note the similar magnitudes
of heavy quark and decay electron quenching if the quark $p_T$ is
rescaled by a factor of $\sim 2$.

In Fig.~\ref{fig:eRAA}, the sensitivity of the  electron quenching to
variations in the heavy quark fragmentation scheme
is shown by the difference between the solid and dashed curves.
The solid curves are calculated as in Ref.~\cite{Cacciari:2005rk} while 
the dashed curves arise when Peterson fragmentation
($\epsilon_c = 0.06$, $\epsilon_b = 0.006$) is used. 
While there can be considerable differences in the fragmentation schemes on
an absolute scale, see Fig.~\ref{fig:ec_eb}, these differences mostly cancel
in the nuclear modification factors shown in Fig.~\ref{fig:eRAA}.

The yellow band corresponding to the combined
$c+b\rightarrow e$ electron sources shows that,
in the kinematic range $4< p_T(e)< 10$ GeV accessible at RHIC,
$R_{AA}(e)$ is dominated by $b$
quark quenching.
Even for the highest opacity, shown on the right-hand side,
we therefore predict that {due to the $b\rightarrow e$ contribution}
\begin{equation}
{R_{AA}(e)>0.5 \; \; {\rm for}\; p_T<6\;{\rm GeV}}
\;\; .
\end{equation}
Increasing the opacity further is not an option
within the theory of radiative energy loss
because pion quenching would then 
be over-predicted.

{The robustness of the bottom
dominance in the electron spectrum
can be seen in the ratio of charm relative to bottom decays to electrons in
Fig.~\ref{fig:ec_eb}.  We use the NLO MNR code \cite{MNR} to compute heavy
quark production for a range of mass and scale values: $1.2 < m_c < 1.7$ GeV,
$4.5 < m_b < 5$ GeV and combinations of the renormalization, $\mu_R$, and
factorization, $\mu_F$, scales such that $(\mu_R/m_T,\mu_F/m_T) =
(1,1)$, (2,1), (1,2) and (2,2).  We employ the same $(\mu_R/m_T,\mu_F/m_T)$
combinations 
for both charm and bottom to maintain the asymptotic approach of the 
distributions at high $p_T$. In all cases, the bottom contribution becomes
larger for $p_T < 5.5$ GeV, even before energy loss is applied.
Changing the Peterson function parameter, $\epsilon_Q$, 
from the standard values
of $\epsilon_c = 0.06$ and $\epsilon_b = 0.006$ to the more delta-function like
values of $\epsilon_c = \epsilon_b = 10^{-5}$ shifts the cross over to
higher $p_T$, more similar to the results with the FONLL fragmentation scheme.
No reasonable variations of the parameters controlling
fragmentation of heavy quarks 
can make the bottom contribution to electrons negligible
at RHIC.}

\begin{figure*}[bht] 
\epsfig{file=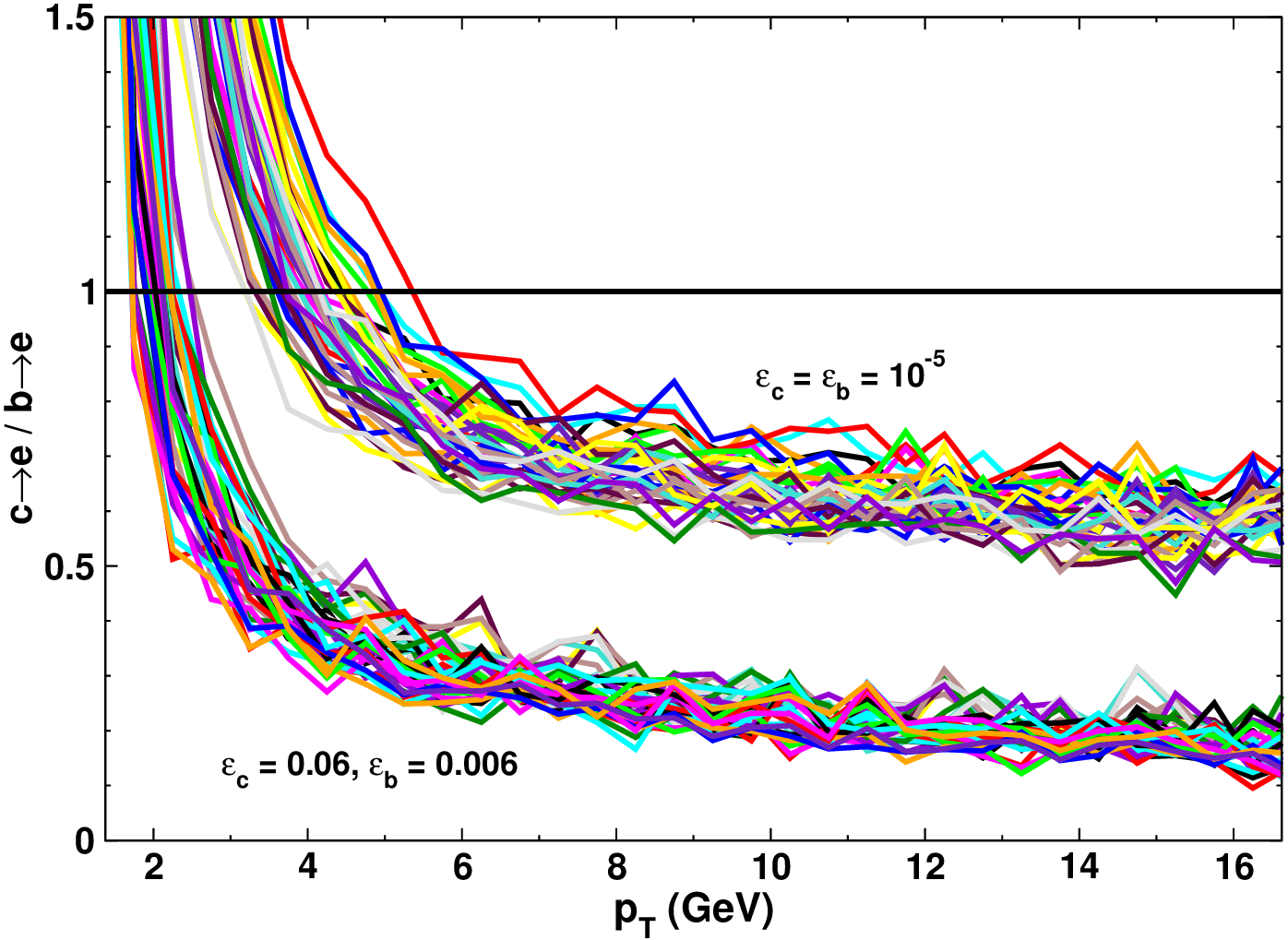,height=2.5in,width=3.0in,clip=5,angle=0} 
\hspace*{0.5cm } 
\epsfig{file=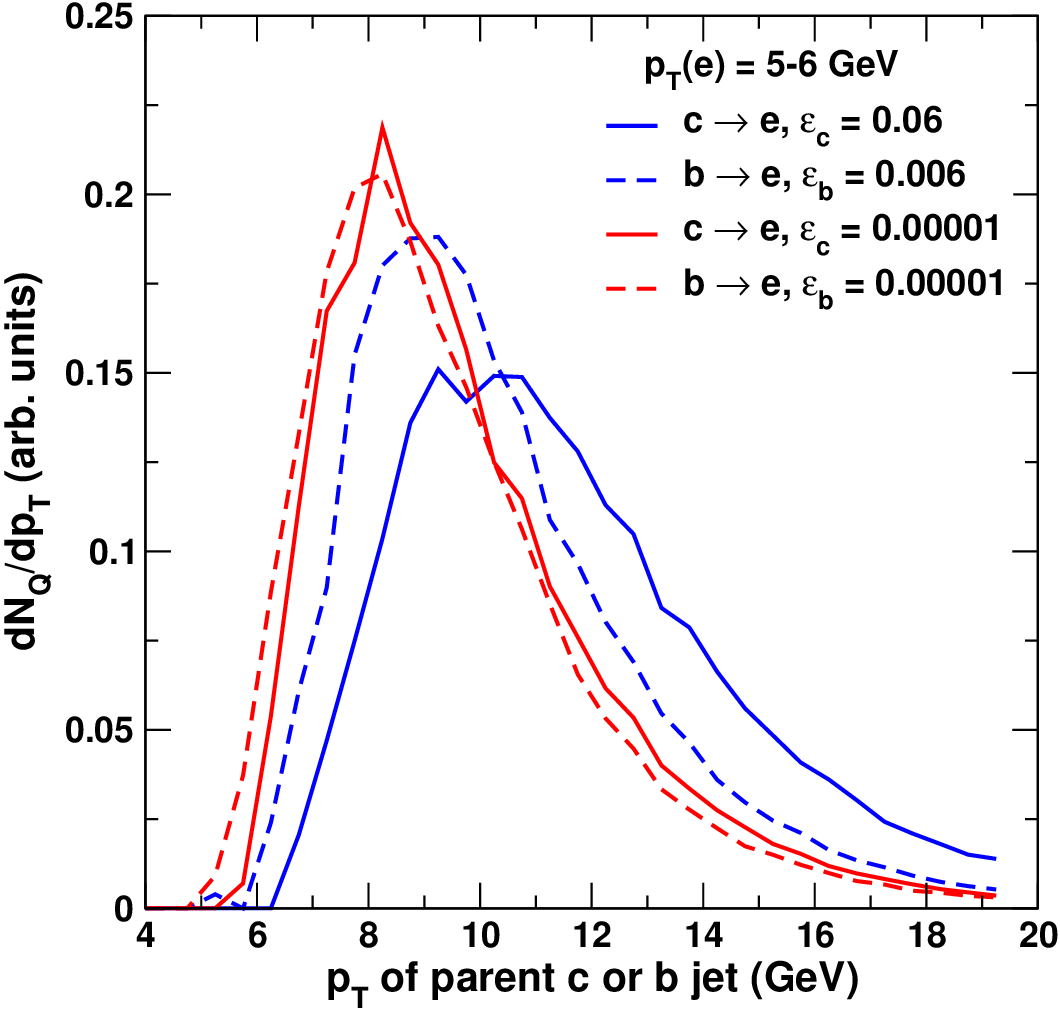,height=2.5in,width=2.5in,clip=5,angle=0} 
\begin{minipage}[t]{15.cm}  
\caption[a]{\label{fig:ec_eb} {[left]} The ratio of
{charm to bottom decays to electrons}
{obtained by varying the quark mass and scale factors.}
The effect
of changing the Peterson function parameters from
$\epsilon_c = 0.06$, $\epsilon_b = 0.006$ (lower band) to 
$\epsilon_c = \epsilon_b = 10^{-5}$ (upper band) is also illustrated.}
\caption[b]{{[right]} \label{fig:eReach}
The electron reach, defined by the distribution 
of initial $c$ and $b$ quark momenta
that after fragmentation and decay
produce an electron with $p_T=5-6$ GeV
 using the PYTHIA fragmentation scheme.
 }
\end{minipage}
\end{figure*}

As a final check, in Fig.~\ref{fig:eReach} we show
the ``electron reach'' defined by 
the transverse momentum distribution of the initial heavy quarks
that decay to electrons with $p_T=5-6$ GeV.
As can be readily seen, this range of electron $p_T$ is sensitive
to heavy quark quenching at approximately twice this scale:
$p_T\sim 6-10$ GeV with hard fragmentation parameters, $\epsilon_c = \epsilon_b
= 10^{-5}$, and $p_T\sim 9-14$ for the standard Peterson parameters.
Given the slow variation of heavy quark quenching in the $p_T\sim 10-20$ 
GeV range seen in Fig.~\ref{fig:qRAA}, it is easy to understand why 
single electron quenching
is robust to uncertainties in the heavy quark fragmentation scheme, as
shown in Fig.~\ref{fig:eRAA}.

{\em Conclusions.} \\
In this letter we predicted the nuclear modification factor
of single electrons,
$R_{AA}(p_T, m_Q, dN_{g}/dy)$,
produced by fragmentation of quenched
{bottom as well as charm} quark jets  
in central Au+Au collisions with $\sqrt{s}=200$ $A$GeV.  
We found that within
the DGLV theory of radiative energy loss,
$b$ quark jets give the dominant contribution
to $p_T \sim 5$ GeV electrons, limiting $R_{AA}(e)>0.5$.
Therefore, if the preliminary PHENIX data suggesting  $R_{AA}(e)<0.5$
are confirmed, it will be a theoretical
challenge to devise novel energy loss mechanisms 
that make the sQGP opaque 
to bottom quarks of $p_T \sim 10-20$ GeV without
over-predicting
the observed light hadron quenching in the $p_T\sim 10$ GeV range.

{\em Acknowledgments:} Valuable discussions with Azfar Adil, Brian Cole,
{John Harris,}
 William Horowitz,
Denes Molnar, Thomas Ullrich, Ivan Vitev and
Nu Xu are gratefully acknowledged.
This 
work is supported by the Director, Office of Science, Office of High Energy 
and Nuclear Physics, Division of Nuclear Physics, of the U.S. Department of 
Energy under Grants No. DE-FG02-93ER40764, DE-AC02-05CH11231.

\end{document}